\documentclass{aa}  
\usepackage{graphicx}
\usepackage{amsmath}
\usepackage{txfonts}
\usepackage{xcolor}
\usepackage{soul}	

\newcommand{\be}{\begin{equation}} 
\newcommand{\ee}{\end{equation}}

\begin{document}

   \title{Cosmic ray protons and electrons from supernova remnants}

   \author{P. Cristofari,
          \inst{1,2,3}\thanks{pierre.cristofari@obspm.fr}
          P. Blasi\inst{2,3}\thanks{pasquale.blasi@gssi.it}
          \and D. Caprioli\inst{4}\thanks{caprioli@uchicago.edu}
          }

   \institute{ 
   \inst{1}Observatoire de Paris, PSL Research University, LUTH, 5 Place J. Janssen, 92195 Meudon, France \\
    \inst{2}Gran Sasso Science Institute, via F. Crispi 7--67100, L'Aquila, Italy\\
   \inst{3}INFN/Laboratori Nazionali del Gran Sasso, via G. Acitelli 22, Assergi (AQ), Italy \\ 
      \inst{4}Department of Astronomy and Astrophysics, University of Chicago, 5640 S Ellis Ave, Chicago, IL 60637, USA 
 }

   \date{Received September 15, 1996; accepted March 16, 1997}

 
  \abstract
   {The spectrum of cosmic ray protons and electrons released by supernova remnants throughout their evolution is poorly known, because of the difficulty in accounting for particle escape and confinement in the downstream of a shock front, where both adiabatic and radiative losses are present. Since electrons lose energy mainly through synchrotron losses, it is natural to ask whether the spectrum released into the interstellar medium may be different from that of their hadronic counterpart. In a rather independent way, studies of cosmic ray transport through the Galaxy require that the source spectrum of electrons and protons be very different, hence the question above also acquires a rather phenomenological relevance.}
   {Here we calculate the spectrum of cosmic ray protons released during the evolution of supernovae of different types, accounting for the escape from upstream and for adiabatic losses of particles advected downstream of the shock and liberated at later times. The same calculation is carried out for electrons, where in addition to adiabatic losses we take into account the radiative losses suffered behind the shock. The latter are dominated by synchrotron losses in the magnetic field that most likely is self-generated by cosmic rays accelerated at the shock.}
   {We use standard temporal evolution relations for supernova shocks expanding in different types of interstellar medium together with an analytic description of particle acceleration and magnetic field amplification to determine the density and spectrum of cosmic ray particles. Their evolution in time is derived by solving numerically the equation describing advection with adiabatic and radiative losses, for electrons and protons. The flux from particles continuously escaping the SNRs is also accounted for.}
   {The magnetic field in the post-shock region is calculated by using an analytic treatment of the magnetic field amplification due to non--resonant and resonant streaming instability and their saturation. The resulting field is compared with the available set of observational results concerning the dependence of the magnetic field strength upon shock velocity. We find that  when the field is the result of the growth of the cosmic-ray--driven non--resonant instability alone, the spectrum of electrons and protons released by a supernova remnant are indeed different, but such a difference becomes appreciable only at energies $\gtrsim 100-1000$ GeV, while observations of the electron spectrum require such a difference to be present at energies as low as $\sim 10$ GeV. An effect at such low energies requires  substantial magnetic field amplification in the late stages of the supernova remnant evolution (shock velocity $\ll 1000$ km/s), perhaps not due to streaming instability but hydrodynamical processes. We comment on the feasibility of such conditions and speculate on the possibility that the difference in spectral shape between electrons and protons may reflect either some unknown acceleration effect, or additional energy losses in cocoons around the sources.}
   {}


   \maketitle

\section{Introduction}

Although there is strong evidence that particle acceleration takes place in supernova remnants (SNRs), it is still debated whether these objects can be the sources of all of the Galactic cosmic rays (CRs).There are different levels of the problem: first, the processes of particle acceleration and magnetic field amplification in an individual SNR depend on the type of explosion and the type of environment where it occurs. Second, the spectrum of particles accelerated at the shock and the one that the SNR releases in the surrounding interstellar medium (ISM) are, in general, quite different. The latter is typically made of two components, the one that escapes the remnant at any given time from the upstream region, and the one that is advected downstream and is eventually liberated when the SNR dissipates. The particles trapped in the downstream are affected by losses, which in general act differently on protons and electrons. Even this simple line of thought leads to two conclusions: 1) although the instantaneous spectrum of particles accelerated at the SNR shock is a power law in momentum, the released spectrum does not need to be so; 2) the spectra of protons and electrons from a SNR can, in general, be different. It is worth keeping in mind that non--linear effects might lead to a slight deviation from the perfect power laws predicted in linear theory~\citep{reynolds1992,malkov2001}.

As previously discussed by \cite{cristofari2020}, the spectrum of CR protons from different types of SNRs is hardly a pure power law and it extends to a maximum momentum that depends rather critically on the type of SNR and on the environment in which the explosion takes place. The structure in the spectrum is due to the contribution of different times during the evolution of the remnant, and the overlap of the advected and escaped fluxes. For type Ia SNe the proton spectrum extends to $\lesssim 100$ TeV, but it is characterized by a marked steepening at $\sim$ TeV energies. Such steepening is associated with the transition between the spectrum of advected particles and that of escaped CRs. The maximum energy at any given time was calculated using the growth rate of the non-resonant hybrid instability \cite[]{bell2004}. For core collapse SNe, in principle higher maximum energies can be reached, but the steepening at the transition discussed above is so pronounced that in fact the flux at the Earth gets strongly suppressed at a few TeV energies. For very energetic ($\gtrsim 5 \times 10^{51}$ erg) SN explosions taking place in a dense pre-supernova wind (called here type II* SNe), having an estimated rate of occurrence of the order of a few per $10^{4}$ years, the normalisation of the flux approaches the measured CR flux (at all energies) and the maximum energy is around the knee. The spectral shape of CRs contributed by each of these SN types is characterized by structures that appear qualitatively similar to the bumps recently observed by DAMPE \citep{DAMPE2019}. 

In the present work we focus on the description of magnetic field amplification and its implications for the maximum energy reached by protons and electrons and radiative losses of electrons trapped in the downstream plasma of a SNR. In particular, we provide a detailed description of the adiabatic losses and escape of protons from SNRs of different types, and of the transport of electrons subject to synchrotron losses. This last calculation has been recently presented by \citet{rebecca} for a typical remnant from a type Ia progenitor, expanding in a uniform ISM,  and the authors concluded that the spectrum of electrons is systematically steeper than that of protons, in line with the requirement arising from calculations of the transport of nuclei and electrons in the Galaxy \citep{EvoliPRL}. However the finding of \citet{rebecca} was based on a recipe for magnetic field amplification derived by \citet{amatoblasi} for resonant streaming instability (but without the natural saturation to $\delta B\sim B$, appropriate for these modes) and modified with a phenomenological recipe, so that the Alfv\'enic Mach number is replaced by the same quantity calculated in the amplified field;
such a prescription was used, e.g., to reproduce the multi-wavelength emission from Tycho's SNR  \citep{morlino2012, slane2014}. This prescription leads to relatively large magnetic field amplification at late times, and maximum energy of electrons that remains loss dominated even for old SNRs. We show that the difference in spectral shape between protons and electrons is very sensitive to the strength of the magnetic field in such late phases of the SNR evolution, and the effect virtually disappears if magnetic field amplification is described solely by Bell instability. We carry out this calculation for three types of SNRs, for both protons and electrons, in order to assess the role of the environment around SNRs for the shaping of CR spectra. 

The article is organised as follows: in \S \ref{sec:Bfield} we summarize our understanding of CR induced magnetic field amplification at SNR shocks and compare the predicted magnetic field with a compilation of observational results for a number of SNRs. In \S \ref{sec:spectra} we describe how we follow protons and electrons in the downstream region of the shock, and we summarize the description of the escape of these particles from upstream. In \S \ref{sec:SNR} we briefly summarize our treatment of the temporal evolution of different types of SNRs in the surrounding environment. In \S \ref{sec:results} we illustrate the main result of our calculations in terms of injection spectra of protons and electrons from different types of SNRs, integrated in time over the whole temporal evolution of the SNR shock through the surrounding medium. In \S \ref{sec:discuss} we discuss the implications of our results. 

\section{Magnetic field amplification in SNRs}
\label{sec:Bfield}

In this section we summarize our current understanding of CR induced magnetic field amplification at SNR shocks. Magnetic field perturbations can be produced in the shock proximity due to a variety of processes, but only some of them lead to important effects in terms of particle scattering. For instance the propagation of a shock front in a medium with density inhomogeneities can excite a Richtmeier-Meshkov instability \cite[]{jokipii2007} that leads to the growth of perturbations downstream of the shock, on a time scale of order $\ell/v_{A}$, if $\ell$ is the spatial scale of the density inhomogeneities upstream and $v_{\rm A}$ the Alfv\'en speed. Although such magnetic field may be important in terms of determining the morphology of synchrotron emission from a remnant  and the strength of the downstream magnetic field, it does not  appreciably affect the diffusion time of accelerated particles upstream, hence it does not  lead to a substantial increase of the maximum energy that can be reached through diffusive shock acceleration (DSA). A fundamental step forward in the investigation of the interaction between CRs and the surrounding medium has been made with the discovery of the non-resonant hybrid instability \cite[]{bell2004}, that is expected to be excited upstream of a shock due to the accelerated particles themselves. This is a current driven instability, excited both by CR particles that are leaving the acceleration region as well as by CRs diffusively confined in the vicinity of the shock front. If the density of CRs with momentum $>p$ at the shock is $n_{\rm CR}(>p)$, the electric current that these particles carry is $J_{\rm CR}(>p)=e D (\partial n_{\rm CR}/\partial z)_{\rm shock}=e v_{\rm sh}n_{\rm CR}(>p)$ and it extends over a precursor distance $\sim D(p)/v_{\rm sh}$. On the other hand, the density of CRs escaping toward upstream infinity is limited to the highest energy particles and can be estimated as $\sim n_{\rm CR}(>p)(v_{\rm sh}/c)$. Hence the corresponding current is $J_{\rm CR}(>p)\approx e n_{\rm CR}(v_{\rm sh}/c) c= e n_{\rm CR} v_{\rm sh}$, numerically equivalent to the one estimated above for the same momentum, despite the fact that escaping particles are assumed to be streaming away ballistically (namely moving at roughly the speed of light, $c$). If the differential spectrum of accelerated particles at the shock is $f_{\rm CR}(p)=A \left( \frac{p}{m_{\rm p}c}\right)^{-\alpha}$, the normalisation can be easily found by requiring that the CR pressure is a fraction $\xi_{\rm CR}$ of the ram pressure at the shock location, $\rho v_{\rm sh}^{2}$:
\be
\frac{1}{3} \int_{p_{\rm min}}^{p_{\rm max}} \text{d}p 4\pi p^{2} f_{\rm CR} (p) p v(p) = \xi_{\rm CR} \rho v_{\rm sh}^{2},
\ee
which implies that $A=(3/4\pi)\xi_{\rm CR}\rho v_{\rm sh}^{2}/(m_{\rm p}^{4}c^{5}I(\alpha))$, with $I(\alpha)=\int_{p_{\rm min}/m_{\rm p}c}^{p_{\rm max}/m_{\rm p}c}\text{d}x ~x^{4-\alpha}/(1+x^{2})^{1/2}$. Notice that the normalisation constant defined in this way depends very weakly on the minimum and maximum momenta $p_{\rm min}$ and $p_{\rm max}$, provided $4\le\alpha <5$, as expected for particle acceleration by DSA. In particular, for $\alpha=4$ one has $I(\alpha)\approx \ln(p_{\rm max}/m_{\rm p}c)$. 
 The spectrum of accelerated particles may be outside this range only if: 1) non-linear effects due to the CR pressure lead to the formation of a precursor upstream, which in turn may lead to spectra harder than $p^{-4}$ \citep[e.g.,][]{jones2001,malkov2001}. However, in practice, strong spectral modification should not be expected because of numerous other effects \cite[]{berezhko1999,2009MNRAS.395..895C}. Moreover, recent self-consistent kinetic simulations of strong shocks suggest that the formation of a shock postcursor naturally leads to spectra with $\alpha$ between 4 and 5 \citep[][]{caprioli+20}, consistent with $\gamma$-ray observations of SNRs \citep[][]{caprioli11}.
In general, the slope depends on the shock Mach number (if the shock is not strong) but even at the end of the Sedov-Taylor phase, the Mach number remains much larger than unity~(see \S \ref{sec:SNR}).


The current carried by accelerated particles with momentum $>p$ is 
$$
J_{\rm CR}(>p)=e v_{\rm sh} \int_{p}^{p_{\rm max}} 4\pi p^{2}A \left( \frac{p}{m_{\rm p}c}\right)^{-\alpha} \approx 
$$
\be
~~~~~~~~~~~~~~~~\approx \frac{3 e v_{\rm sh}}{m_{\rm p}c^{2}} \frac{\xi_{\rm CR} \rho v_{\rm sh}^{2}}{(\alpha-3) I(\alpha)}\left( \frac{p}{m_{\rm p}c}\right)^{3-\alpha}.
\ee
Here we especially focus on the scenario where the non-resonant hybrid instability is excited by escaping particles, since this channel leads to the formation of magnetic perturbations far upstream. As discussed by~\cite{bell2004}, the fastest growing mode is associated with a wavenumber $k_{\rm max}$ and can be written as $\gamma_{\rm max}=k_{\rm max}v_{\rm A}$, where $v_{\rm A}=B_{0}/\sqrt{4\pi \rho}$ is the Alfv\'en speed in the unperturbed field $B_{0}$. The wavenumber $k_{\rm max}$ is determined based on the condition:
\be
k_{\rm max} B_{0} \approx \frac{4\pi}{c} J_{\rm CR}(>p) ,
\ee
and the excitation of the instability occurs only if $k_{\rm max}>r_{\rm L}(p)=pc/eB_{0}$, which translates to the following constraint:
\be
pc~n_{\rm CR}(>p)  \frac{v_{\rm sh}}{c} \gtrsim \frac{B_{0}^{2}}{4\pi}.
\ee
In other words the instability is excited if the energy density in the form of escaping particles is larger than that of the pre-existing magnetic field. As discussed in much previous literature, there are different approaches to the saturation of the instability. The most intuitive one, based on comparing the plasma displacement due to the $J_{\rm CR}\times \delta B/c$ force with the Larmor radius of particles in the amplified field $\delta B$ leads to the conclusion that the field stops growing when the energy density in the form of escaping particles equals that in the amplified field:
\be
\frac{\delta B^{2}}{4\pi}\approx 3\frac{v_{\rm sh}}{c}\frac{\xi_{\rm CR}\rho v_{\rm sh}^{2}}{(\alpha-3)I(\alpha)}\left(\frac{p}{m_{\rm p}c}\right)^{4-\alpha}.
\label{eq:satura}
\ee
At saturation, the spatial size of the perturbations becomes comparable with the Larmor radius in the amplified field, so that the current is now disrupted because of efficient CR scattering off the self-generated perturbations, thereby causing the drive for magnetic field amplification to stop. 

Some comments on Eq.~\eqref{eq:satura} are in order: 1) from the point of view of scattering of particles with momentum $p$, the diffusion coefficient is $D(p)\propto p/\delta B$. If the spectrum of accelerated particles is $\sim p^{-4}$ then the diffusion coefficient is Bohm-like (linear in momentum). In a general case, $\alpha=4+\epsilon$, the diffusion coefficient turns out to be $D(p)\sim p^{1+\epsilon/2}$. Since typically $\epsilon\sim 0\div 0.3$, the expected deviations from Bohm-like behaviour are small. 2) With the exception of the case $\epsilon=0$, Eq.~\eqref{eq:satura} shows a weak dependence upon the momentum $p$ where the current is calculated. 3) In terms of magnetic field immediately upstream of the shock, all particles should be included in estimating the magnetic field in Eq.~\eqref{eq:satura}. For $\alpha=4$ ($\epsilon=0$) the resulting field does not depend on this choice, but for $\epsilon>0$ the resulting magnetic field shows a weak dependence on the minimum momentum. Assuming a minimum momentum $\chi m_{\rm p} v_{\rm sh}$, with $\chi>1$, one has:
\be
\frac{B_{1}^{2}}{8\pi \rho} \approx \frac{3}{2} \chi^{4-\alpha} \left(\frac{v_{\rm sh}}{c}\right)^{5-\alpha}\frac{\xi_{\rm CR} v_{\rm sh}^{2}}{(\alpha-3)I(\alpha)}.
\label{eq:B1}
\ee
If the turbulent field upstream of the shock is roughly isotropic and the perpendicular components are compressed at the shock, with compression factor $r$, the mean value of the compressed downstream magnetic field is $B_{2}\approx B_{1}\sqrt{(1+2r^{2})/3}$. For $r=4$, the compression factor is $\sqrt{11}$. It follows that:
\be
\frac{B_{2}^{2}}{8\pi \rho} \approx \frac{1}{2}(1+2r^{2})\chi^{4-\alpha} \left(\frac{v_{\rm sh}}{c}\right)^{5-\alpha}\frac{\xi_{\rm CR} v_{\rm sh}^{2}}{(\alpha-3)I(\alpha)}.
\label{eq:B2}
\ee
\begin{figure}[h]
\includegraphics[width=.5\textwidth]{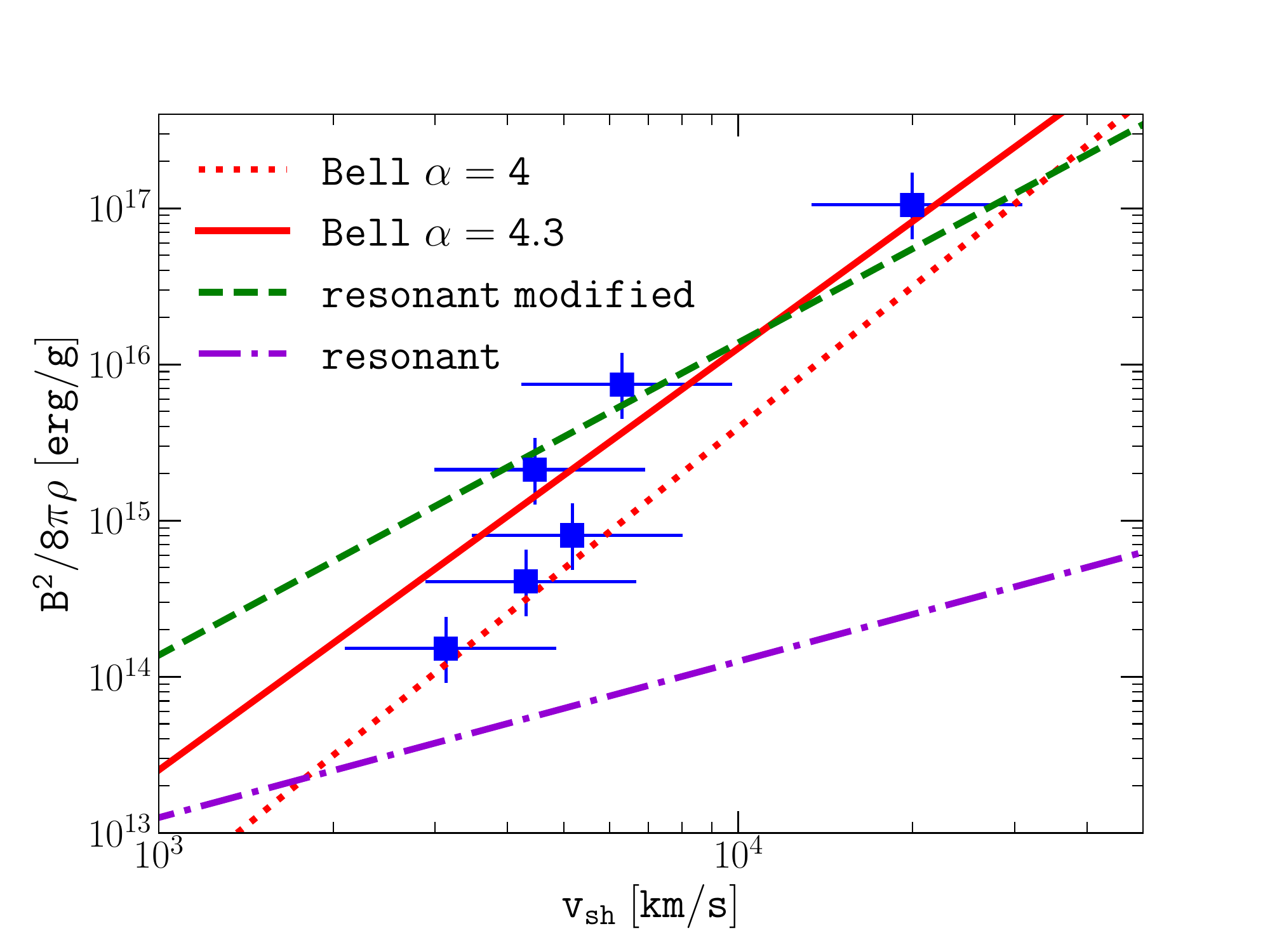}
\caption{Magnetization of the downstream of SNR shocks as a function of shock velocity. The data points are from \cite{vink2012}, while the lines refer to the growth of non-resonant modes with an injection spectrum with slope $\alpha=4$ (solid red) and $\alpha=4.3$ (dotted red) and to the recipe adopted by \cite{rebecca} (dashed green). The latter is independent of the spectrum of accelerated particles.
}
\label{fig:vink}
\end{figure}
 For the parameters' values appropriate to the Tycho SNR, this expression returns a downstream magnetic field that ranges between 150 $\mu$G ($\alpha=4$) and 400 $\mu$G ($\alpha=4.3$), in good agreement with the value inferred from multi-wavelength observations \citep[e.g.][]{morlino2012}.

In Fig. \ref{fig:vink} we compare the quantity in Eq.~\eqref{eq:B2} with the corresponding value as measured in several SNRs, as reported by \cite{vink2012}. The solid (dotted) line shows the result of Eq.~\eqref{eq:B2} for the case $\alpha=4.3$ ($\alpha=4$), assuming $\chi=3$. When the amplified field upstream becomes smaller than $B_{0}$, the magnetic field relevant for synchrotron losses becomes of order $\sqrt{11}B_{0}$. For typical values of parameters in the ISM this reflects in $B_{2}^{2}/8\pi \rho \sim 2\times 10^{12}$ erg/g in Fig. \ref{fig:vink}. This condition also identifies the end of the stage where non resonant modes get excited  \citep[see also][]{kinetic2009}, and typically occurs when the $v_{sh}\lesssim 1000$ km/s. 

Eq.~\eqref{eq:satura} clearly shows how, for this type of instability,  when saturation is reached the quantity $\delta B^{2}/8 \pi \rho$ scales with the shock velocity as $\propto v_{\rm sh}^{3}$ ($\propto v_{\rm sh}^{2.7}$ for $\alpha=4.3 $) and is independent of the strength $B_{0}$ of the pre-existing magnetic field. Other recipes for saturation (see for instance the one proposed by~\citet{riquelme} suggest some weak dependence of the magnetic field at saturation on $B_{0}$. These recipes lead to somewhat lower amplified magnetic field, a point to keep in mind in the discussion below. 

The dashed (green) curve in Fig.~\ref{fig:vink} represents the same quantity calculated following the recipe originally put forward by \citet{caprioli11} and used in \cite{morlino2012}.
A few comments are in order concerning the genesis of this approach. Resonant streaming instability was originally included in non-linear theories of DSA by \cite{amatoblasi}, where the equation for the growth of these modes was solved analytically. The saturation level derived from this simple approach can be written as:
\be
\frac{B_{1}^{2}}{8\pi \rho} \approx \frac{1}{4M_{\rm A}}\xi_{\rm CR} v_{\rm sh}^{2},
\label{eq:ab06}
\ee
where $M_{\rm A}=v_{\rm sh}/v_{\rm A,0}$ is the Alfv\'en Mach number calculated with respect to the Alfv\'en speed in the field $B_{0}$, $v_{\rm A,0}=B_{0}/\sqrt{4\pi\rho}$.  It follows that the compressed field downstream reads:
\be
\frac{B_{2}^{2}}{8\pi \rho} \approx \frac{1+2r^{2}}{12M_{\rm A}}\xi_{\rm CR} v_{\rm sh}^{2}.
\label{eq:ab06compr}
\ee
It is worth noting that, since $M_{A}=v_{\rm sh}/v_{\rm A,0}$, effectively $B_{2}^{2}\propto v_{sh}$. 
In Fig. \ref{fig:vink}, this is shown as a dashed purple line. 
We should stress that this amplified field should be used only when the Bell modes are not allowed to grow (low shock velocity). In the opposite condition the growth rate of the resonant instability is proportional to $n_{CR}^{1/2}(>p)$ and leads to a lower magnetic field than the non resonant instability, and usually $\delta B/B_{0}\lesssim 1$~\citep{zweibel1979,kinetic2009}.



\cite{morlino2012} proposed a formal modification of Eq.~\eqref{eq:ab06compr}, so as to mimic the onset of non-linear effects: the proposed modification consists in attempting to interpret the Mach number in Eq.~\eqref{eq:ab06} as the one calculated with respect to the Alfv\'en speed in the amplified field $\delta B$. This assumption results in a modified expression for the amplified field that, after compression at the shock, reads:
\be
\frac{B_{2}^{2}}{8\pi \rho}\approx \frac{1+2r^{2}}{24} v_{\rm sh}^{2} \xi_{\rm CR}^{2}.
\label{eq:B2res}
\ee
This quantity is plotted in Fig. \ref{fig:vink} for $r=4$ (dashed green line). The most distinctive characteristic of this trend is the scaling with $v_{\rm sh}^{2}$, quite different from the $\propto v_{\rm sh}^{3}$ typical of non--resonant hybrid modes (see Eq.~\eqref{eq:B2} for $\alpha=4$). Eq.~\eqref{eq:B2} results in a much larger magnetic energy density downstream at late times, when the shock velocity drops down. 

 Though being a viable phenomenological approach, this is based on a few assumptions that is worth having in mind: 1) although being based on a perturbative approach, Eq. \ref{eq:ab06} leads to $B_{1}/B_{0}\gg 1$ for high shock speed, which is unphysical if modes are strictly resonant; 
2) as pointed out above, the growth rate used to describe resonant modes and leading to Eq.~\eqref{eq:ab06}, is only valid in the case in which the energy density of particles is much smaller than $B_{0}^{2}/4\pi$, which is certainly not the case for fast shocks (in fact this is the very regime where non--resonant hybrid modes grow). This was discussed at length by~\citet{kinetic2009} \citep[see also][]{review2013}.

To add to this rather confused situation, hybrid kinetic simulations of particle acceleration at shocks lead to a prescription for the amplified field that reads \citep{caprioli2014a,caprioli2014b}:
\be
\frac{B_{1}^{2}}{B_{0}^{2}} \approx 3 M_{A} \xi_{CR},
\label{eq:pic}
\ee
formally similar to Eq. \ref{eq:ab06} but with a different numerical factor. It is not easy to encapsulate these simulation results in the theory above: the simulations were ran for a range of parameters in which Bell modes would grow and they are seen to be growing. On the other hand, these simulations are non relativistic, which implies that the anisotropy of the accelerated particles (which for an astrophysical shock is $\sim v_{\rm sh}/c\sim 10^{-2}\div 10^{-3}$) is $\sim v_{\rm sh}/v\sim 0.1$. This might lead to a larger fraction of the CR energy being channelled into CRs compared with a SNR shock. Moreover, if accelerated particles are forced to remain non relativistic, the growth of the Bell instability can be shown to saturate to a value that is larger than the one in Eq.~\eqref{eq:B1}. 
In any case no dependence upon the background field $B_{0}$ is  expected, based on semi-analytical arguments, unless more complex phenomena come into play. 
Finally, the magnetic field energy density in Eq. \ref{eq:pic} shows a scaling $B_{2}^{2}/\rho \propto v_{sh}$, but the non-relativistic simulations discussed by \cite{caprioli2014a,caprioli2014b} were ran for a given shock velocity and changing $B_{0}$ so as to achieve different Alfv\'en Mach numbers. 
This procedure is optimal to unravel the scaling of $B^{2}$ with $v_{\rm sh}/v_A$, but would not reveal the additional scaling with the actual CR speed, $c/v_A$. These aspects definitely deserve further investigation.

 In the following we focus on the investigation of the two prescriptions for magnetic field amplification described in Eq. \eqref{eq:B2} and Eq. \eqref{eq:B2res}, but we also comment on the case in which the field is amplified through non-resonant streaming instability when the appropriate condition is satisfied and by resonant streaming instability (Eq. \eqref{eq:ab06}) at later times. 

A quick inspection of Fig. \ref{fig:vink} shows that data points are too sparse to allow us to infer a clear dependence of the amplified field on shock speed $v_{\rm sh}$, although there is a mild preference for a $\propto v_{\rm sh}^{3}$ trend. As one can see from Eq.~\eqref{eq:B2}, the non--resonant hybrid modes are expected to lead to $B_{2}^{2}/8\pi\rho\sim v_{\rm sh}^{7-\alpha}$, which compares well with the observed trend for $\alpha=4\div 4.3$. As discussed below, most of the modification of the electron spectrum due to energy losses occurs at late times (low shock speeds). Hence, for the purpose of calculating the difference in spectrum between electrons and protons, it is of critical importance to understand what is the downstream magnetic field for older SNRs. 

The importance of the non--resonant hybrid instability for particle scattering is well known and will be briefly summarised here only for the sake of completeness. Following~\citet{2013MNRAS.431..415B} and~\citet{cristofari2020}, the maximum energy of protons can be  estimated by requiring the condition $\int_0^{t} \text{d}t' \gamma_{\rm max}(t') \simeq 5$, which leads to:
\be
\left( \frac{p_{\rm max}}{m_{\rm p}c}\right)^{\alpha-3} = \frac{3e R_{\rm sh}}{10 m_{\rm p}c^{2}}\frac{\sqrt{4\pi \rho}}{c}
\frac{\xi_{\rm CR}v_{\rm sh}^{2}}{(\alpha-3)I(\alpha)}.
\label{eq:Emax}
\ee
Introducing the expression for the total magnetic field upstream, $B_{1}$, one can rewrite this expression as
\be
\left( \frac{p_{\rm max}}{m_{\rm p}c}\right)^{\alpha-3} = \frac{4}{10} \left(\frac{\tilde\Omega_{\rm c} R_{\rm sh}}{c}\right) \chi^{\alpha-4} 
\left( \frac{v_{\rm sh}}{c}\right)^{\alpha-4} 
\left( \frac{\tilde v_{\rm A}}{v_{\rm sh}}\right),
\ee
where $\tilde\Omega_{\rm c}=e B_{1}/m_{\rm p}c$ is the cyclotron frequency and $\tilde v_{\rm A}=B_{1}/\sqrt{4\pi \rho}$ is the Alfv\'en speed, both calculated in the amplified magnetic field $B_{1}$.  It is worth pointing out that for $\alpha=4.3$ the total magnetic field at the shock is slightly larger than for $\alpha=4$, but the corresponding $p_{\rm max}$ is lower, because there is less power available at the scales resonant with particles with momentum $p_{\rm max}$.

The maximum momentum of electrons is computed by equating the acceleration time to the minimum between the synchrotron time and the age of the SNR~\citep[see e.g.][]{blasi2010}.

\section{Cumulative spectra of CRs at SNRs}
\label{sec:spectra}
The contribution of CRs from an individual SNR can be written as the sum of two components: 1) particles accelerated at the SNR shock and trapped downstream of the expanding shell, until the time when the shock dissipates away and the particles are released into the ISM after having suffered the effect of losses, $N_{\rm loss}$, and 2) particles accelerated at the shock up to the highest energy that can be achieved at that time and leaving the acceleration region from upstream, $N_{\rm esc}$. Although the latter contribution is strongly peaked around the maximum energy reached at that time, the integration over the whole expansion history of the remnant leads to a continuous spectrum of CRs released into the ISM \citep{caprioli2009}. 

The spectrum of particles accelerated at a strong shock in the test--particle limit~\citep{caprioli2009}, assuming a free escape boundary condition at some location upstream, reads: 
\begin{equation}
\label{eq:f0}
f^{\rm p}(p,t)= A(t) \exp \left[ - \frac{3 r }{r -1} \int_{p_{\rm inj}}^{p} \frac{\textrm{d}p'}{p'} \frac{1}{1- \exp \left[- \frac{p_{\rm max}(t)}{p'} \right]}  \right].
\end{equation}
For $p \ll p_{\rm max}(t)$, $f^{\rm p}(p) \propto (p/p_{\rm inj})^{-3r/(r-1)}$, and $p \gg p_{\rm max}(t)$, $f^{\rm p}(p) \propto  \exp \left[- \frac{p}{p_{\rm max}(t)}\right]$.

Working under the usual assumption that a fraction of the ram pressure of the SNR shock is converted into CRs, it is easy to write the spectrum of CRs accelerated at the SNR shock at any given time. In a given time interval $\textrm{d}t$, the number of particles of momentum $p$ $\textrm{d}n_{\rm acc}(p,t)$ accelerated at the shock reads: 
\begin{equation}
4\pi p^{2} \text{d}n_{\rm acc}(p,t) = \textrm{d}t 4 \pi R_{\rm sh}^2(t) \;  v_{\rm sh}(t)/r \; f^{\rm p}(p,t) 4\pi p^{2},
\label{eq:dnacc}
\end{equation}
where $v_{\rm sh}/r$ is the velocity downstream of the SNR shock and $r$ is the compression factor at the shock. 

If accelerated particles could escape the acceleration region immediately after penetrating downstream, the total number of particles integrated over the entire \textit{active} lifetime of a SNR, from $t_{0}$, typically the beginning of the free expansion phase, to $T_{\rm SN}$, the end of the Sedov--Taylor (ST) phase, would read: 
\begin{equation}
\label{eq:Nacc}
4\pi p^{2}N_{\rm acc}(p) = \int_{t_{0}}^{\rm T_{\rm SN}} \textrm{d}n_{\rm acc}(p,t) 4\pi p^{2}. 
\end{equation}
This quantity could be interpreted as the flux of CR particles contributed by a SNR in the absence of adiabatic energy losses. The departures from such spectrum will be used later as an index of the effect of losses on the spectrum of CRs from each SNR. In principle, if the SNR shell is broken in some locations where escape is allowed, the time integrated CR flux would be somewhat similar to that estimated using Eq. \ref{eq:Nacc}.

The spectrum of electrons at the SNR shock is calculated as in~\citet{morlino2009}, using the approximated expression proposed by~\citet{zirakashvili2007}: 
\begin{equation}
f^{\rm e} (p,t) = K_{\rm ep} f^{\rm p}(p,t) \left[1 + 0.523 \left(  \frac{p}{p^{\rm e}_{\rm max} (t)}  \right)^{9/4} \right]^2 \exp \left[  -  \left( \frac{p}{p^{\rm e}_{\rm max} (t)} \right)^2\right].
\end{equation}
 where K$_{\rm ep}$ is the electron--to--proton ratio, typically in the range $10^{-4} - 10^{-2}$.
The cumulative spectrum of electrons $N^{\rm e}_{\rm acc}$ can then be calculated as in Eq.~\eqref{eq:Nacc}, using $f^{\rm e}$, instead of  $f^{\rm p}$.

Below, we calculate the spectrum of protons and electrons that leave a SNR after accounting for the effect of adiabatic and radiative energy losses. 

\subsection{Adiabatic and radiative losses of CRs downstream of the shock}

As the SNR shock expands, the particles produced at the shock, and trapped inside the SNR downstream of the shock, suffer adiabatic and radiative losses. The latter are dominated by the emission of synchrotron photons, while inverse Compton scattering,  although included in our calculation,  is typically negligible.  In fact, the rate of inverse-Compton losses on cosmic microwave background photons is the same as that of synchrotron losses in a $\sim 3\mu G$ magnetic field, and post-shock fields are much larger due to amplification and compression. On the other hand, Inverse-Compton losses might become important for SNRs located in star-forming regions, if the energy density of optical/infra-red photons exceeds $\sim 100$ eV/cm$^{3}$.

The number of particles (protons or electrons) that are liberated by an individual SNR at the end of the evolution can be easily written in terms of conservation of the total number of particles. In fact, the number of particles with momentum $p$ at the end of the SN evolution ($t=T_{\rm SN}$) is the result of all the particles produced at earlier times ($t<T_{\rm SN}$) with momentum $p'>p$, such that in a time $T_{SN}$ the momentum has degraded down to $p$. We can then write:
\begin{equation}
\label{eq:number}
N^{\rm p,e}_{\rm loss}(p) = \int_{t_0}^{T_{\rm SN}}  \textrm{d}t \frac{4 \pi}{r} R_{\rm sh}^2(t) \;  v_{\rm sh}(t) \left(\frac{p'}{p}\right)^2\; f^{\rm p,e}(p',t) \frac{\text{d}p'}{\text{d}p}. 
\end{equation}

The change of momentum of a particle injected at a time $t'$ with momentum $p'$ due to losses is: 
 \begin{equation}
\frac{\text{d}p}{\text{d}t}= - \frac{p}{\cal L} \frac{\text{d} \cal L}{\text{d}t} + \frac{4}{3} \sigma_{\rm T} c \left( \frac{p}{m_{\rm e} c} \right)^2 \frac{B_2^2(t)}{8 \pi},
\label{eq:change}
\end{equation}
where $\sigma_{\rm T}$ is the Thompson cross--section and ${\cal L}$ accounts for adiabatic energy losses, in terms of change of volume between the two times $t'$ and $t$: 
 \begin{equation}
{\cal L} (t,t')= \left( \frac{\rho_{\rm down}(t)}{\rho_{\rm down}(t')}\right)^{1/3},
\label{eq:conservation}
\end{equation}
where$\rho_{\rm down}$ the density downstream of the shock. If the expansion is adiabatic, $\rho_{\rm down}\propto P^{1/\gamma} \propto (\rho v_{\rm sh}^2(t))^{1/\gamma}$, and $\rho$ is the gas density upstream of the shock). For protons, synchrotron losses are negligible while for electrons both adiabatic and radiative losses are important.

\subsection{Escaping particles}

The flux of particles escaping the accelerator from the upstream region can be written following~\citet{caprioli2009} (see also \citet{cristofari2020}):
\begin{equation}
N^{\rm p,e}_{\rm esc}(p)= \int_{t_{0}}^{\rm T_{\rm SN}}  \textrm{d}t' \frac{4 \pi }{r} r^2_{\rm sh}(t') v_{\rm sh}(t')  f^{\rm p,e}(p,t')  G(p, t') ,
\label{eq:q_acc}
\end{equation}
where we introduced the function 
\begin{equation}
G(p, t')= \frac{\exp \left[-\frac{p_{\rm max}(t')}{p}\right]}{1- \exp\left[ - \frac{p_{\rm max}(t')}{p}\right]},
\end{equation}
that describes the spectral shape of the particles that can escape from upstream at a given time. For protons, the function $G(p,t)$ is strongly peaked around the maximum momentum of protons $p_{\rm max}(t)$ at the given time $t$. For electrons, when the maximum energy is determined by diffusion, the meaning of this function is the same as for protons. When the magnetic field is large enough that synchrotron losses dominate the maximum energy of accelerated electrons (namely for young SNRs), the function $G$ is vanishingly small and no escape from upstream is possible.

\section{Evolution of the shock in the circum--stellar environment}
\label{sec:SNR}

The evolution in time of SNR shocks directly impacts the spectra of particles injected by these sources into the ISM, when integrated on the SNR life span. In this section we briefly summarize the calculations adopted here for the description of such evolution, discussed in more detail in~\citet{cristofari2013}.

For the sake of calculating the contribution to the CR flux, SNe are broadly classified in two groups, depending on the mechanism triggering the explosion: thermonuclear SNe (type Ia) and core--collapse SNe (type II). In addition, we consider a peculiar type of very energetic core--collapse SNe, called here type II*. This type is introduced here just to illustrate the wide range of physical parameters that apply to type II SNe, and to demonstrate that these energetic events can accelerate particles up to the PeV range (when $E_{\rm SN}$ and $\dot{M}$ are sufficiently high, and $M_{\rm ej}$ is sufficiently low, see discussion on the parameter space presented by~\citet{cristofari2020}). 
Schematically, SNR shocks from type Ia SNe expand in a uniform ISM.  The evolution of the shock radius $R_{\rm sh}$ and velocity $v_{\rm sh}$ in the ejecta dominated phase and ST phase are well described by self--similar solutions~\citep[see e.g.][]{chevalier1982,truelove1999,ptuskin2005}. Here we rely on the approach presented in~\citet{truelove1999,truelove2000}, adopting the same formalism as in~\citet{cardillo2015}: 
\begin{equation}
\begin{aligned}
R_{\rm sh} (t)= R_0 \left[ \left(\frac{t}{t_0}\right)^{a \lambda_{\rm FE}}   + \left(\frac{t}{t_0}\right)^{a \lambda_{\rm ST}}\right]^{-1/a}  \\
v_{\rm sh} (t)= \frac{R_0}{t_0} \left(\frac{R}{R_0}\right)^{1-a} \left[ \lambda_{\rm FE}  \left(\frac{t}{t_0}\right)^{a \lambda_{\rm FE}-1}   +  \lambda_{\rm ST} \left(\frac{t}{t_0}\right)^{a \lambda_{\rm ST}-1}\right]^{-1/a} 
\end{aligned}
\end{equation}
where $a=-5$, $R_0= \left( \frac{3 M_{\rm ej} m}{4 \pi n_0}\right)^{1/3}$, $\lambda_{\rm FE}=4/7 $, $\lambda_{\rm ST}= 2/5$, and $t_0= \left[ R_0 \left( \frac{m n_0 M_{\rm ej}}{ 0.38 E_{\rm SN}^2}\right)^{1/7} \right]^{7/4}$. This description holds until the end of the ST phase (i.e. beginning of the radiative phase) of the SNR evolution. This transition typically occur when the age of the SNR becomes of the order of the cooling time: $t_{\rm cool} \approx10^{3} \left( \frac{n_0}{1 \text{cm}^{-3}}\right)^{-1} \left( \frac{v_{\rm sh}(t)}{10^8 \; \text{cm/s}}\right)^3$~kyr~\citep{blondin1998}. In this work, we assume as reference values for type Ia SNe $E_{\rm SN}= 10^{51}$~erg, $M_{\rm ej}=1$~M$_{\odot}$, $n_0=1$~cm$^{-3}$. It is worth recalling that in the presence of efficient CR acceleration at the SNR shock, as discussed by \cite{2018PhRvL.121i1101D}, the evolution in time of the shock in the final stages of the evolution may be affected rather remarkably by the CR pressure and the beginning of the radiative phase may be delayed. We do not include these effects here. 

In general, type II SNR shocks expand in a complex medium structured by the evolution of the massive progenitor star before the explosion of the SN. Throughout its main sequence, the stellar wind produces a low density hot temperature bubble, in pressure equilibrium with the  ISM. Later, when the massive star reaches the final stages of stellar evolution, typically entering the red super giant stage (RSG), a slow dense wind forms. Therefore, after the SN explosion, the shock expands through a dense wind, then through a low density bubble, and finally reaches the ISM. The density of the dense wind created by the RSG is typically $n_{\rm w}= \dot{M}/(4 \pi m u_{\rm w}r^2)$, with $\dot{M} \sim 10^{-5}$ M$_{\odot}$/yr is the mass--loss rate, $u_w\sim 10^{6}$ cm/s the velocity of the wind, and $m=m_p (1+4 f_{\rm He})/(1+ f_{\rm He}) \sim 1.27 m_p$ is the mean mass of the ISM per hydrogen nucleus~\citep{weaver1977}. The density of the low density bubble is $n_{\rm b}= 0.01 (L_{36}^6 n_0^{19} t_{\rm Myr}^{-22})^{1/35}$ cm$^{-3}$, with $t_{\rm Myr}$ is the duration of the main sequence, of the order of $\sim$ Myr~\citep{longair1994}. The transition between the dense RSG wind and the low density cavity, $r_1$ is set by equating the RSG wind pressure to the  thermal pressure of the hot cavity: $r_1= \sqrt{\dot{M}u_{\rm w}/(4 \pi k n_{\rm b} T_{\rm b})}$ where $k$ is the Boltzmann's constant. The radius of the hot bubble is $r_{\rm b}= 27 (L_{36}/1.27 n_0)^{1/5} t_{\rm Mpc}^{3/5}$~pc, where $L_{36}$ is the main sequence star wind power in units of $10^{36}$ erg/s. 
In the case of type II SNe, the evolution in the structured medium described above does not allow for direct self--similar solutions. It is however possible to work under the \textit{thin--shell} approximation, considering that the swept--up gas is located in a think layer behind the shock wave~\citep[see e.g.][]{ostriker1988,bisnovatyikogan1995}. In the case of spherically symmetric distribution, \citet{ptuskin2005} obtained: 
\begin{equation}
\begin{aligned}
v_{\rm sh}(R_{\rm sh})= \frac{\gamma +1}{2} \left[ \frac{12 (\gamma-1) E_{\rm SN}}{(\gamma +1)M^2(R_{\rm sh})  R_{\rm sh}^{6(\gamma-1)/(\gamma+1)} } \right.\\
\left. \times \int_0^{R_{\rm sh}} \text{d}r r^{6\left( \frac{\gamma-1}{\gamma+1} \right)-1} M(r)  \right]^{1/2} \\
\end{aligned}
\end{equation}
and: 
\begin{equation}
 t(R_{\rm sh})= \int_0^{R_{\rm sh}} \frac{\text{d}r}{v_{\rm sh} (r)}
\end{equation}

For a typical type II SN, we assume  $E_{\rm SN}= 10^{51}$~erg, $M_{\rm ej}=5$~M$_{\odot}$, $\dot{M}=10^{-5}$~M$_{\odot}$/yr. For, the type II*, $E_{\rm SN}= 5 \times10^{51}$~erg, $M_{\rm ej}=1$~M$_{\odot}$, $\dot{M}=10^{-4}$~M$_{\odot}$/yr. 

\section{Results}
\label{sec:results}

In this section we discuss our results, in terms of spectrum of protons and electrons injected into the ISM by different types of SN explosions. As discussed above, in general the CR spectrum released by an individual source is the sum of the contribution due to particles escaped from upstream at any given time and the contribution of CRs trapped in the downstream, where energy losses are in action, and liberated in the final stages of the SN evolution, that we assume coincide with the beginning of the radiative phase. 
While the proton escape flux is always present, the electron escape flux vanishes when radiative losses limit the maximum energy.

It is this phase of radiative and adiabatic losses that can potentially make the overall spectrum of electrons different from that of protons, if radiative losses are sufficiently severe.

\begin{figure}[h]
\includegraphics[width=.44\textwidth, height=0.40\textwidth]{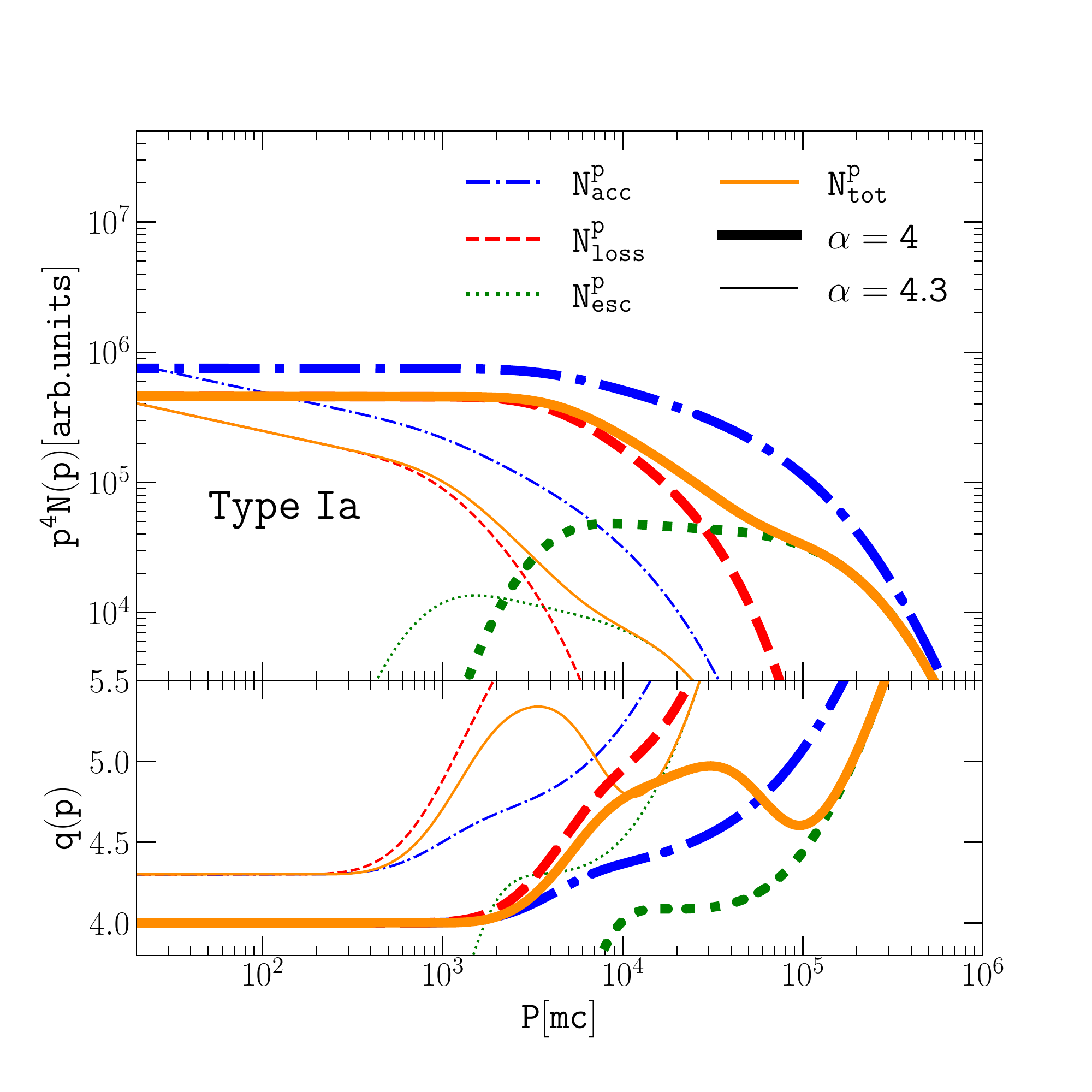}
\includegraphics[width=.44\textwidth,height=0.40\textwidth]{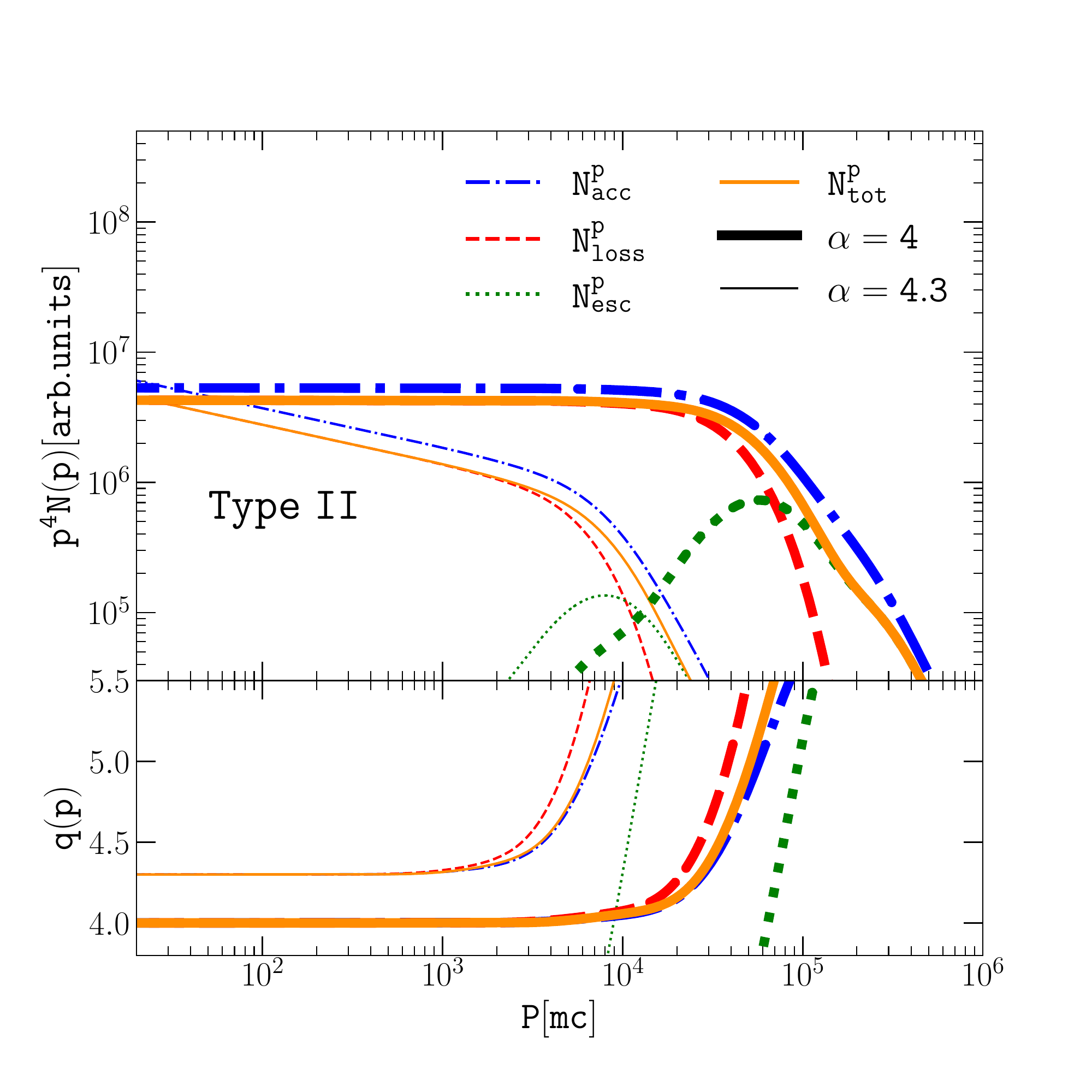}
\includegraphics[width=.44\textwidth,height=0.40\textwidth]{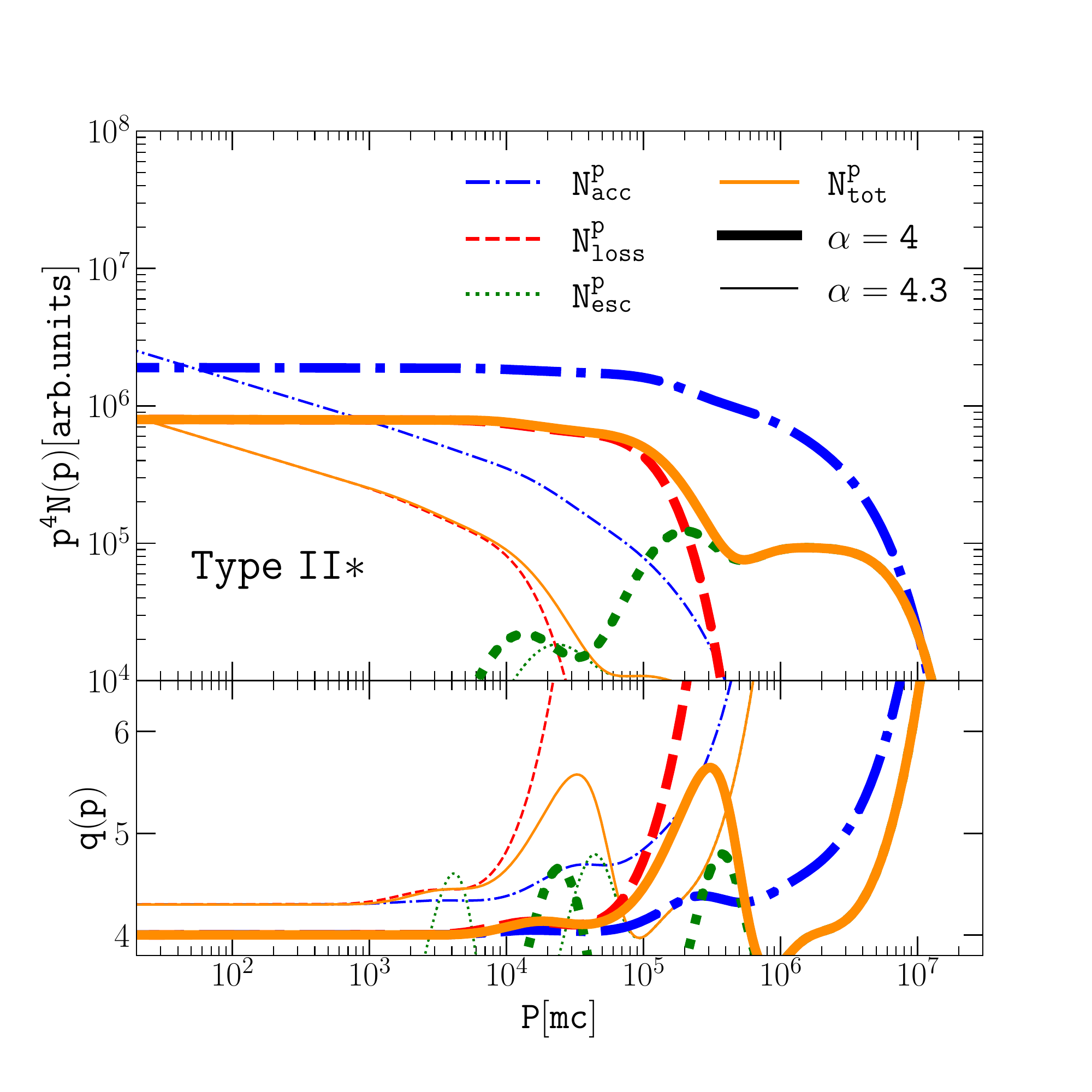}
\caption{Spectra of protons produced at SNRs from type Ia (left), type II (center) and type II* (right) SNRs for $\alpha=4$ (thick lines) and $\alpha=4.3$ (thin lines) if they were instantaneously liberated into the ISM (broken shell assumption). The dashed curves illustrate the effect of adiabatic losses in the downstream region, while the dotted lines refer to the escape flux from upstream. In the bottom part of each panel we also show the local slope of the spectrum $q(p)$ at given momentum.}
\label{fig:protons}
\end{figure}

The spectrum of protons liberated by type Ia, II and II* SNRs is shown in Fig. \ref{fig:protons}, with the corresponding slopes shown in the bottom panels. As discussed by \citet{cristofari2020}, the highest energies are reached at very early times, but they do not reflect in equally large energies in the final spectrum of protons because the amount of mass that the SN shock processes at such early times is very small. In fact the effective maximum energy is the one reached at the beginning of the ST phase. The easiest case is that of type Ia SNRs, where the explosion occurs in the normal ISM, with spatially constant gas density and background magnetic field. For type Ia SNRs the effective maximum energy is a few tens of TeV (left panel of Fig. \ref{fig:protons}). There is an additional spectral steepening at somewhat lower energies due to the temporal evolution of the maximum energy. More specifically the steepening occurs at the maximum energy reached at the end of the ST phase, typically a few TeV. The flux of escaping CR protons starts at about the same energy, as clearly visible in Fig. \ref{fig:protons}. 

For a strong shock, as the one expected for a young SNR expanding in the normal ISM, the spectrum of accelerated particles at the shock location has a slope very close to $4$ (thick lines in Fig. \ref{fig:protons}). Nevertheless, as recently discussed by \citet{caprioli+20}, the spectrum can be steeper if the finite velocity of scattering centres in the downstream plasma is taken into account. For this reason, in Fig. \ref{fig:protons} we also show the case $\alpha=4.3$ as thin lines. In all cases of interest, the spectra of CR protons that are injected into the ISM (as the sum of the two contributions) is quite close to the spectrum at the shock, in terms of slope, with the exception of the highest energies, as discussed above. 

For type II SNRs, the spectrum of CR protons is shown in the middle panel of Fig. \ref{fig:protons}. For the sake of making a fair comparison of the three types of SN explosions, here we used an acceleration efficiency $\xi_{\rm CR}=0.1$ for all of them. As discussed by \citet{cristofari2020}, because of the different rates of occurrence of these events in the Galaxy, for type II SNRs the efficiency is required to be somewhat lower than for type Ia, which also reflects in a lower value of the maximum energy of particles accelerated at the shock (see Eq.~\eqref{eq:Emax}). Despite this bias, the maximum achievable energy for type II SNRs remains of the order of $\sim 10^{5}$ GeV and falls short of the knee by a large amount, as already pointed out by \citet{cristofari2020}. 

Only when parameters are pushed to the extreme (what we called here type II* SNRs) the maximum energy can reach the knee, as shown in the right plot of Fig. \ref{fig:protons}. As already pointed out by \citet{caprioli2009}. The superposition of the escape flux from the different stages of evolution of the shock in the complex environment around these SNRs may lead to the appearance of bumps in the overall CR spectrum that might be related to the feature recently measured by DAMPE in the region 10-100 TeV of the proton spectrum \cite[]{DAMPE2019}.

\begin{figure}[h]
\includegraphics[width=.44\textwidth, height=0.4\textwidth]{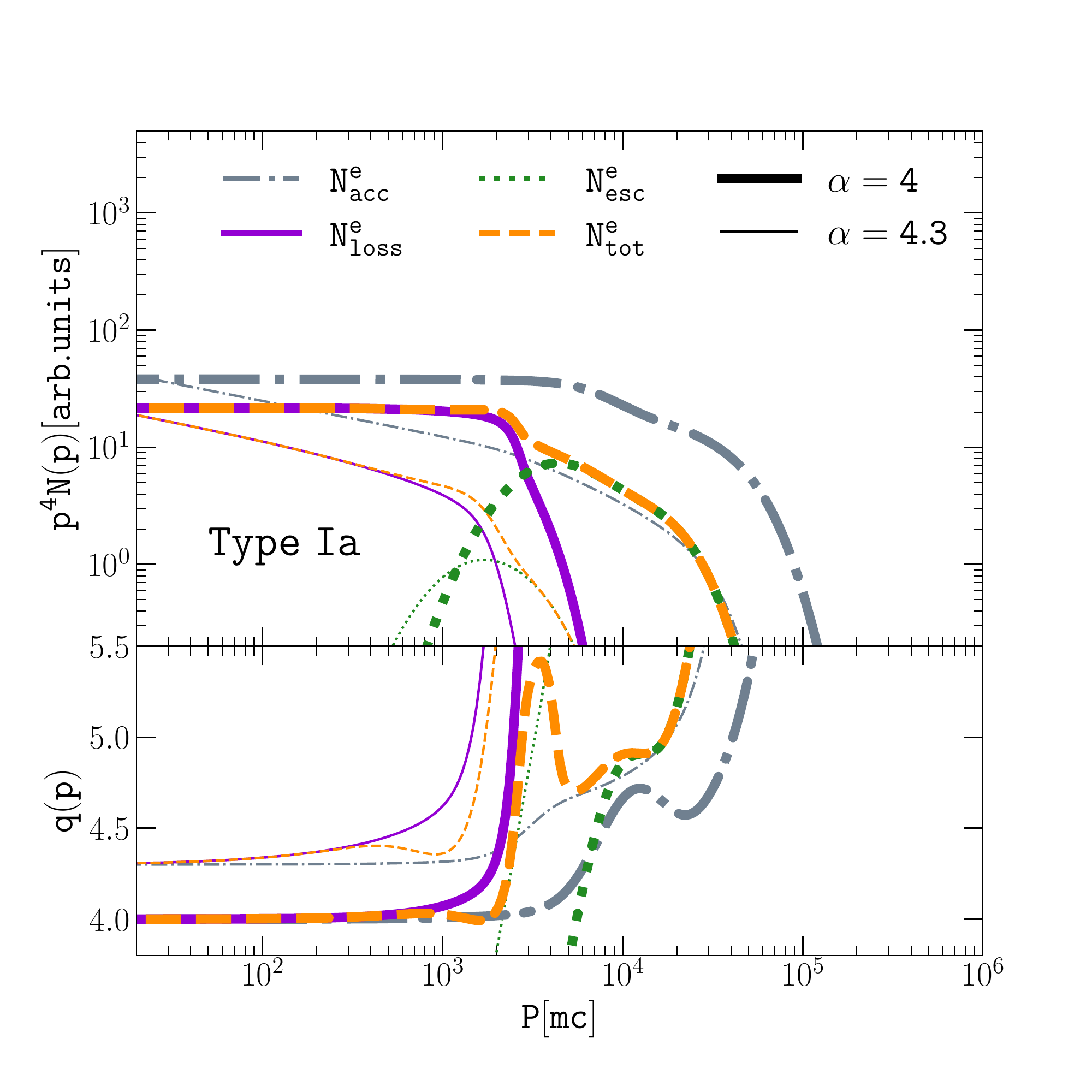}
\includegraphics[width=.44\textwidth, height=0.4\textwidth]{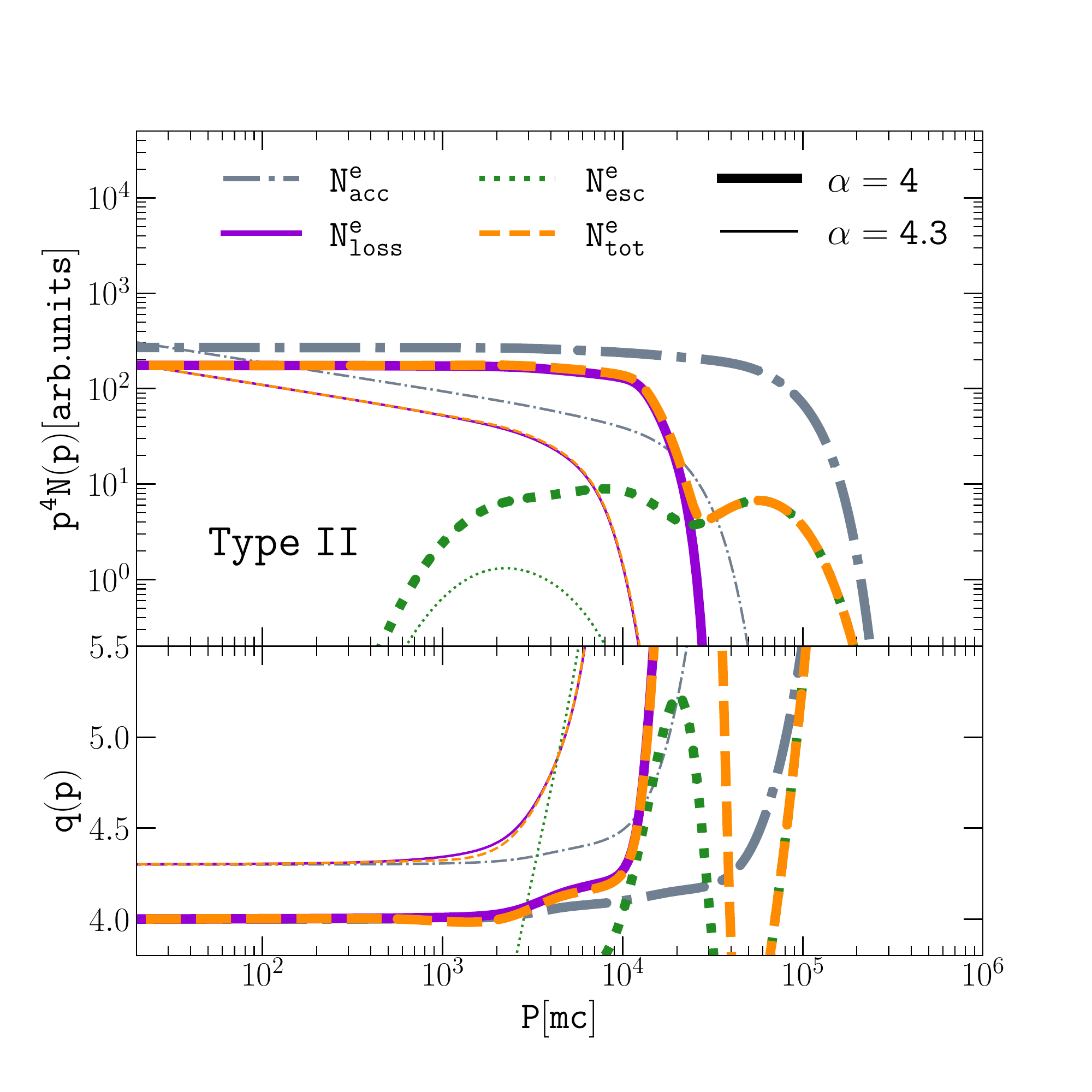}
\includegraphics[width=.44\textwidth, height=0.4\textwidth]{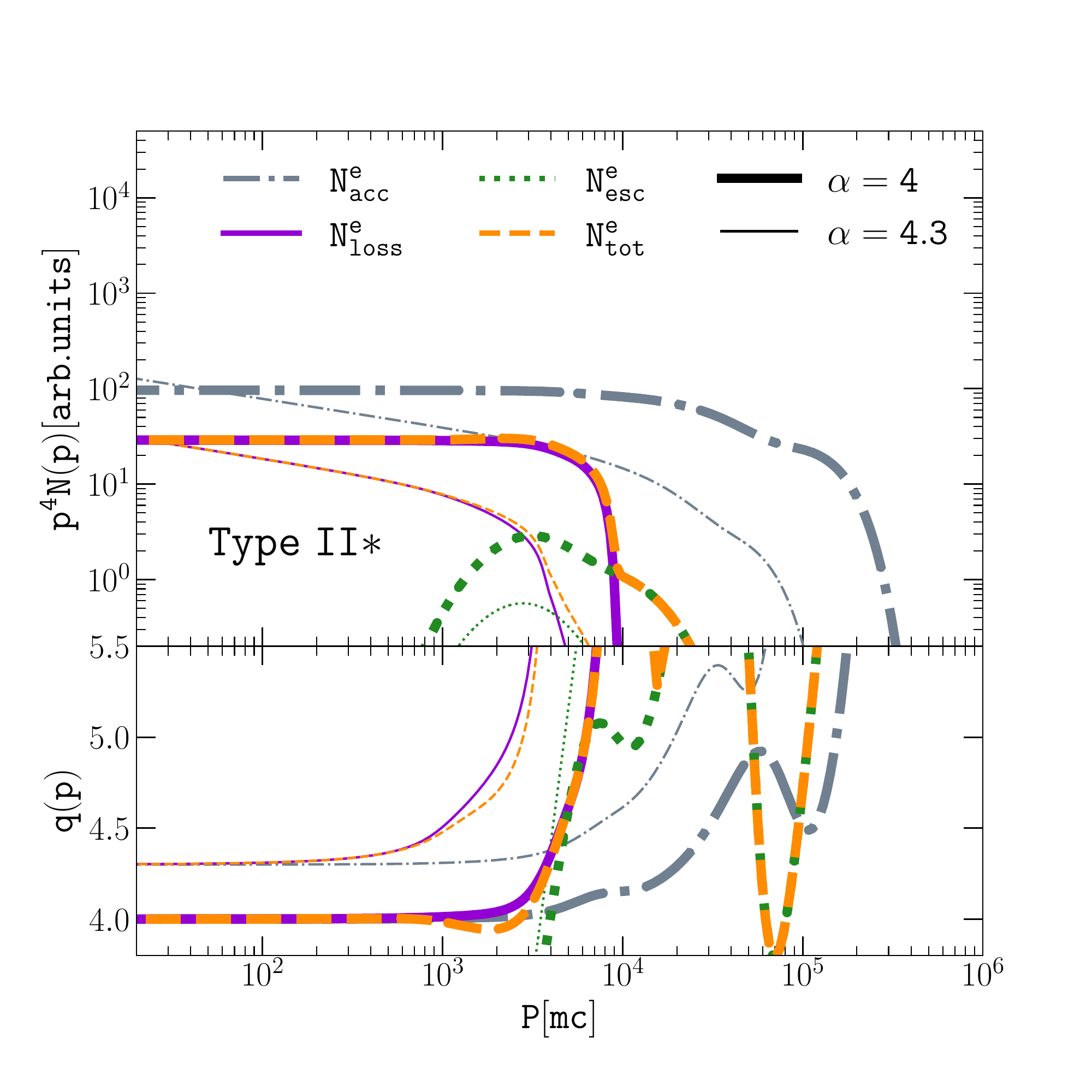}
\caption{Spectra of electrons produced at SNRs from type Ia (left), type II (center) and type II* (right) progenitors for $\alpha=4$ (thick lines) and $\alpha=4.3$ (thin lines). Dash-dotted line: spectrum of particles accelerated at the shock. Solid lines: spectra of electrons liberated into the ISM after losses downstream of the shock. Dotted lines: upstream escape flux. In the bottom part of each panel we also show the slope of the spectrum $q(p)$ at given momentum.}
\label{fig:electrons}
\end{figure}

The corresponding spectra of electrons injected by SNRs of different types into the ISM are shown in Fig. \ref{fig:electrons}. The thick (thin) curves refer to $\alpha=4$ ($\alpha=4.3$) respectively. The dash-dotted line identifies the spectrum of particles accelerated at the shock, as if they were immediately liberated into the ISM, without energy losses. The solid lines are the spectra of electrons liberated into the ISM after adiabatic and synchrotron losses downstream of the shock, while the upstream escape flux, limited to the times when the maximum energy of electrons is not determined by energy losses, is shown in the form of dotted lines.  
If the SNR shell were broken, or confinement in the downstream were energy dependent (e.g., due to turbulence damping) the actual contribution would lie between the dash-dotted and solid lines.

The rate of synchrotron losses is larger when the condition for the growth of magnetic field through the excitation of the non--resonant hybrid instability is fulfilled. As discussed in \S \ref{sec:Bfield}, $B_{2}^{2}/\rho\propto v_{\rm sh}^{7-\alpha}$ for this instability, hence the mechanism becomes less effective or even ineffective in the late stages of a SNR evolution, which are however crucial for the production of low energy electrons. As a consequence, the effect of radiative energy losses is only important at energies $\gtrsim$TeV, while it is minor at lower energies, as clearly shown in Fig. \ref{fig:electrons}, independent of the type of SNR considered. It follows that, if the magnetic field amplification is mainly due to the excitation of Bell modes, then losses cannot be the main reason for a difference in the spectra of protons and electrons in SNRs. 

\begin{figure}[h]
\includegraphics[width=.45\textwidth, height=0.33\textwidth]{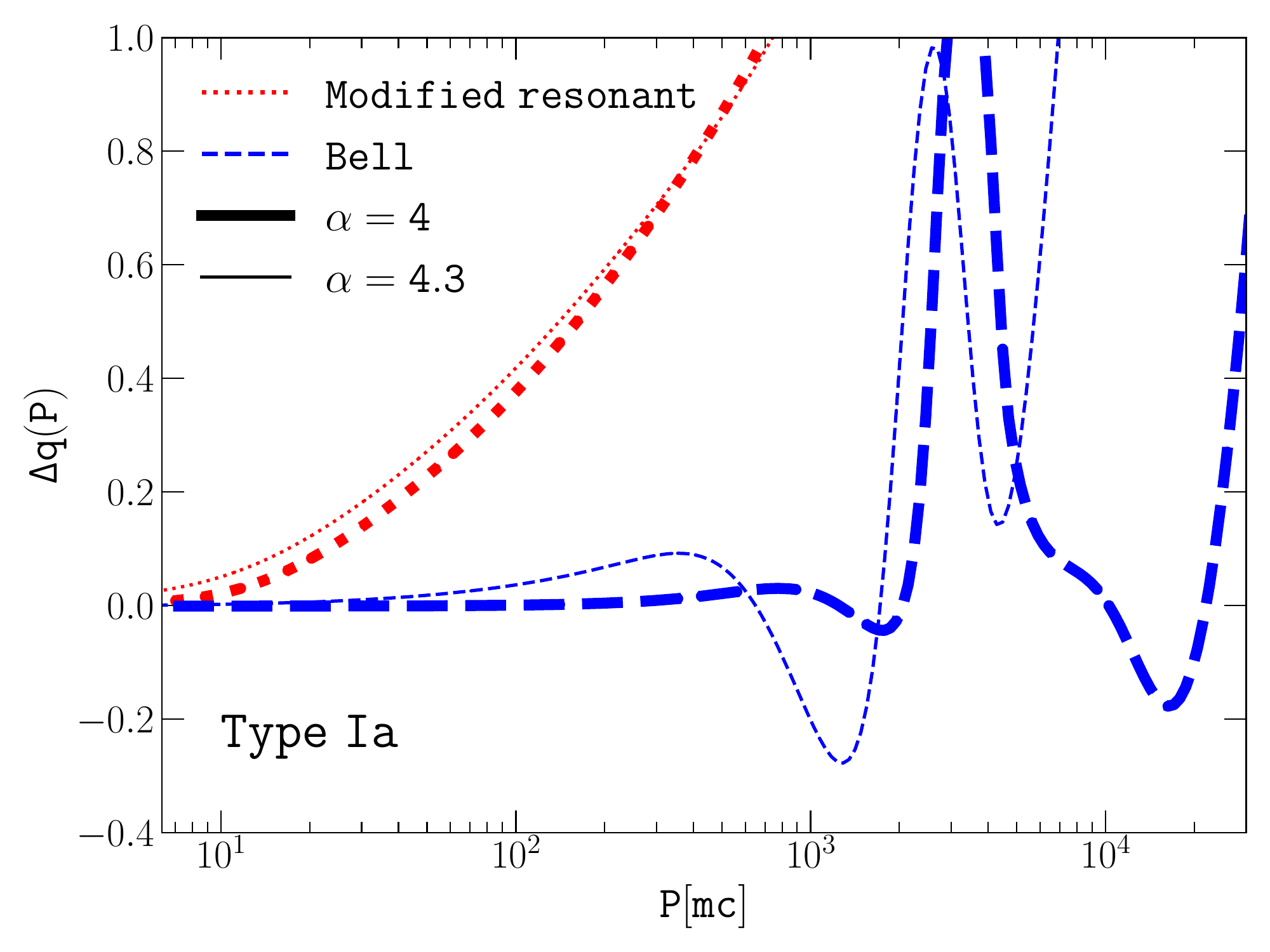}
\includegraphics[width=.45\textwidth, height=0.33\textwidth]{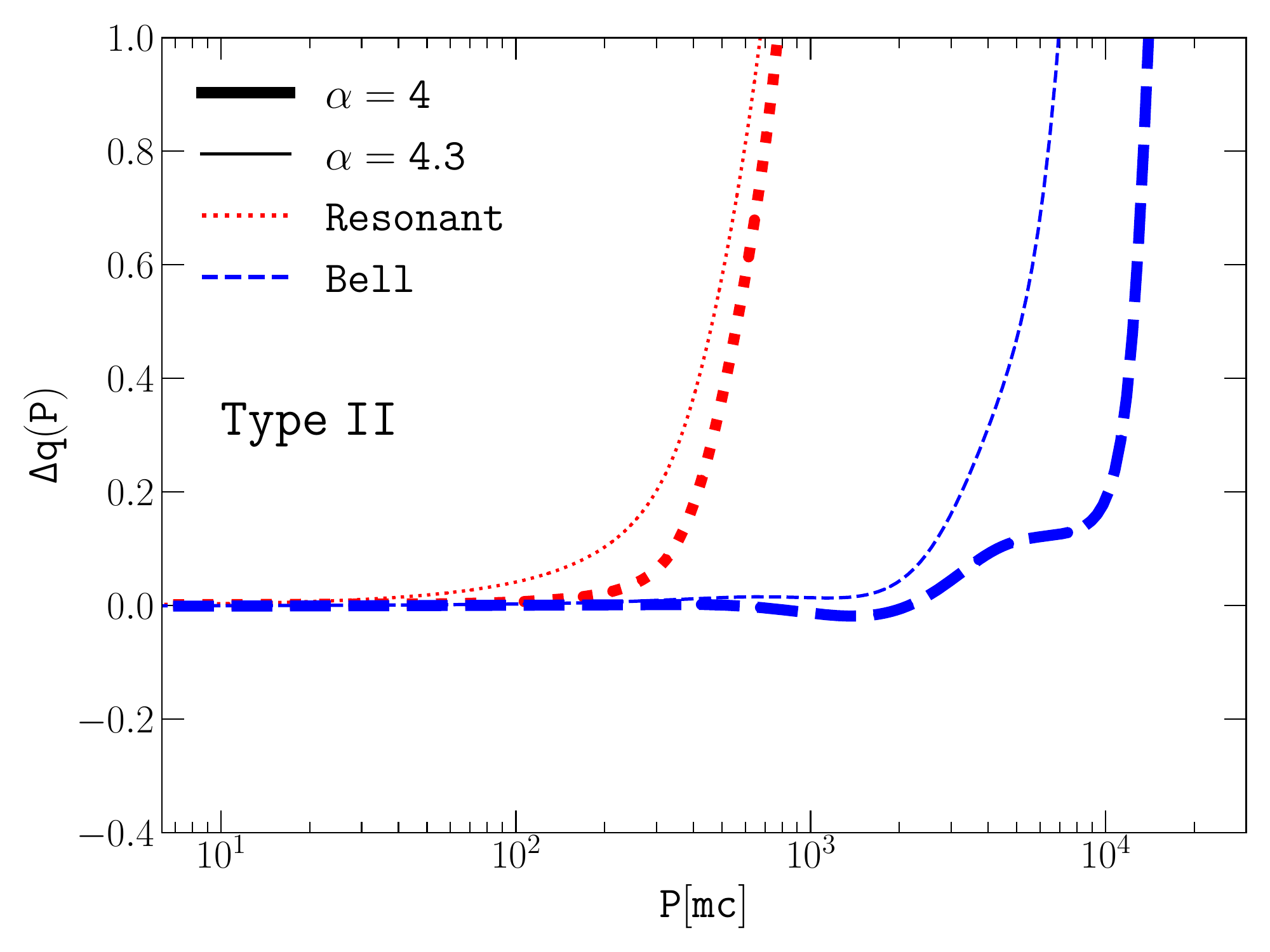}
\includegraphics[width=.45\textwidth, height=0.33\textwidth]{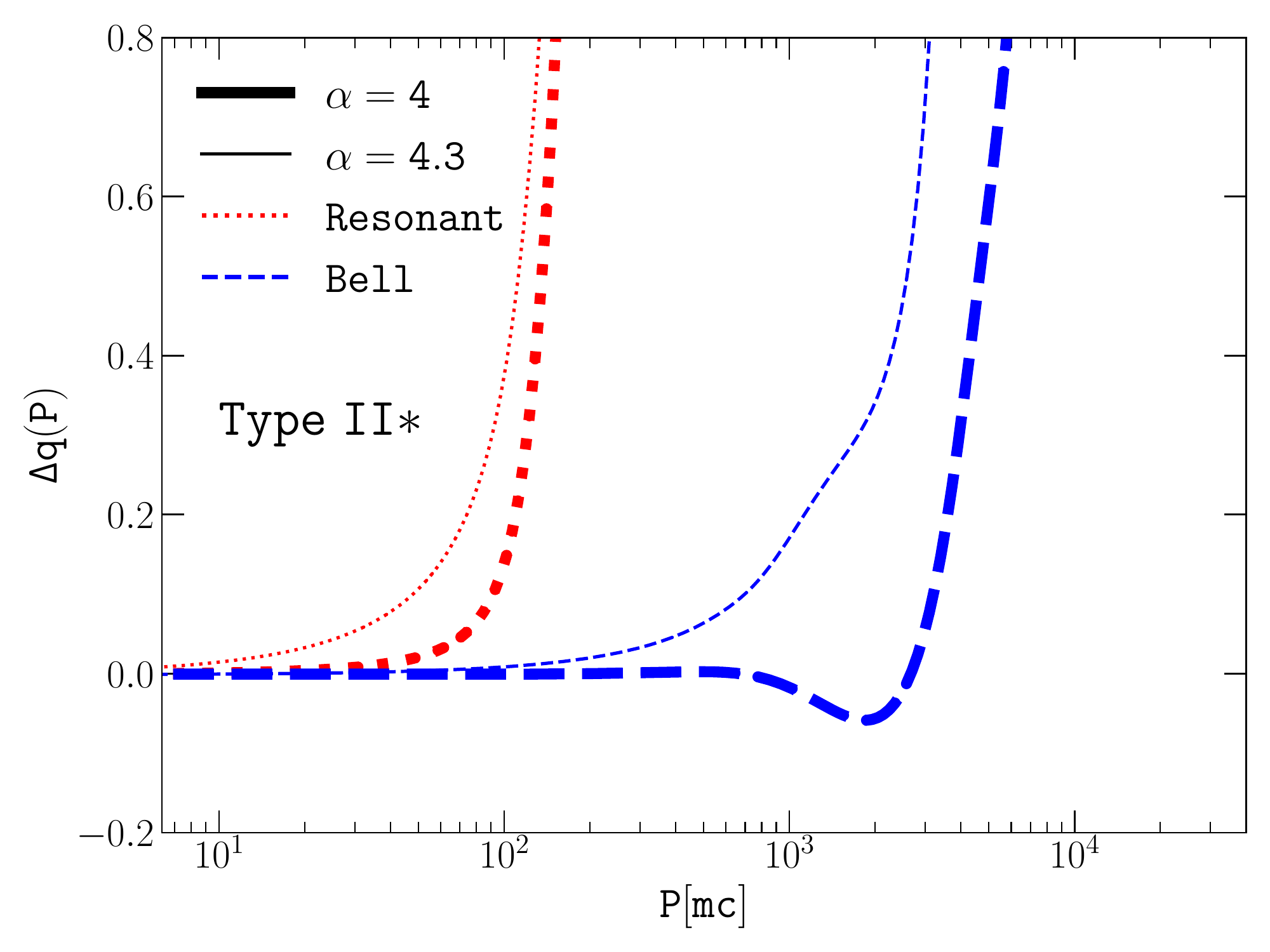}
\caption{Difference between the proton and electron  spectral index at SNRs from type Ia (left), type II (center) and type II* (right) progenitors for $\alpha=4$ (thick lines) and $\alpha=4.3$ (thin lines). Dash-dotted line: spectrum of particles accelerated at the shock. Solid lines: spectra of electrons liberated into the ISM after losses downstream of the shock. Dotted lines: upstream escape flux. In the bottom part of each panel we also show the local slope of the spectrum $q(p)$ at given momentum.}
\label{fig:slopes}
\end{figure}

This result is illustrated even more clearly in Fig. \ref{fig:slopes}, where we show the difference in slope between protons and electrons as a function of energy for type Ia (left), type II (middle) and type II* (right) SNRs. The thick (thin) curves refer to the case $\alpha=4$ ($\alpha=4.3$). The dashed lines are obtained using the growth of non--resonant hybrid modes to calculate the magnetic field, while the dotted lines are based on Eq. \eqref{eq:B2res}, the same as in~\citet[]{rebecca}. The reason for the very different results is that the two recipes lead to quite different predictions for the downstream magnetic field at late times, which are the most important for determining the electrons' spectrum: the non resonant instability leads to intense field in the early phases when the shock is faster, but the instability virtually shuts off at later times, when the shock velocity drops below $\sim 1000$ km/s. On the other hand, the prescription used by~\citet{rebecca} leads to retain larger magnetic fields even in late stages, so that the acceleration of electrons remains loss-dominated at such late times. This reflects in steeper electron spectra. 

The main lesson to be learnt from this exercise is that the spectrum of electrons liberated by SNRs into the ISM is strongly dependent upon the knowledge of the magnetic field in the phases of the SNR evolution where we know the least. 

The issue boils down to whether we have observational bounds or theoretical arguments that can be used to assess the credibility of different scenarios. From the observational point of view, to our knowledge, there is no evidence in favour or against having relatively large magnetic field amplification at late times. From the theoretical point of view, the only prejudice that we can mention is that the recipe based on Eq. \eqref{eq:B2res} is somewhat weak, for the reasons explained above. If to be somewhat more conservative and limit our attention to scenarios that are based on some sort of theoretical grounds, one could say that the magnetic field downstream is provided by Eq. \eqref{eq:B2} when the energy density of accelerated particles satisfies the condition for the growth of the non resonant instability, and by Eq.~\eqref{eq:ab06compr} when the non resonant instability stops growing and only resonant Alfv\'en modes get excited. In both cases, when the amplified magnetic field becomes much smaller than $B_{0}$, the field to be used for the purpose of calculating the electrons' energy losses reduces to $B_{0}$.

\begin{figure}[h]
\includegraphics[width=.5\textwidth]{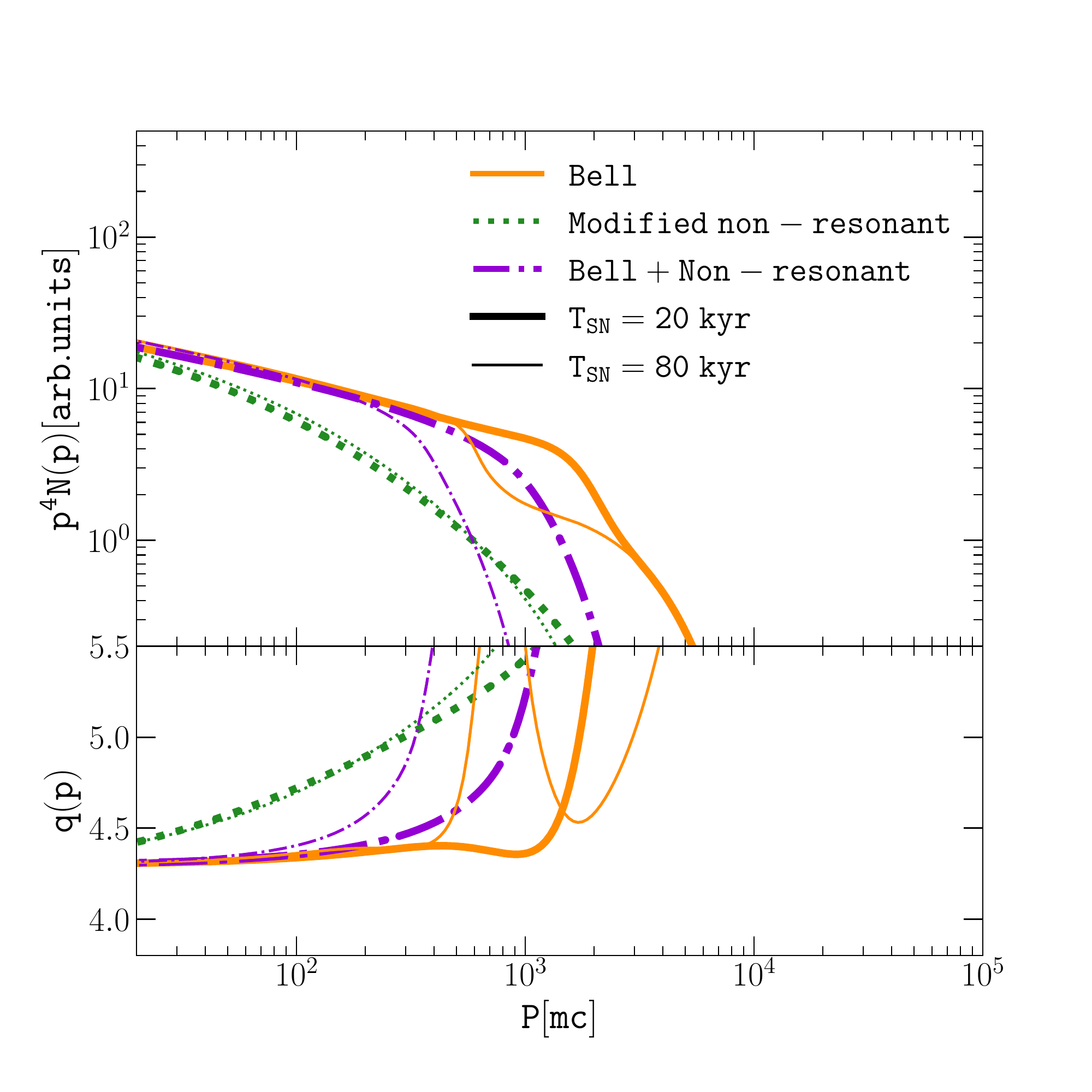}
\caption{Spectra of electrons produced at SNRs from type Ia progenitors for $\alpha=4.3$. We illustrate the effect of the different prescriptions for the magnetic field amplification, and an increased duration of the ST phase, $T_{\rm SN}=80$ kyr (thin lines). Dash-dotted line: spectrum of particles accelerated at the shock. Solid lines: spectra of electrons liberated into the ISM after losses downstream of the shock. Dotted lines: upstream escape flux. In the bottom part of each panel we also show the slope of the spectrum $q(p)$ at given momentum.}
\label{fig:others}
\end{figure}

 To further address the importance of the late times in the SNR evolution  for shaping the electron spectrum, we consider a rather extreme situation in which particle acceleration is efficient until 80 kyr. Indeed, it has been shown that the duration of the ST phase can be substantially extended, for instance because of the CR pressure on the SNR shock \cite[]{2018PhRvL.121i1101D}. In Fig.~\ref{fig:others}, we illustrate the results for a type Ia SN (for a slope 4.3 of the spectrum of accelerated particles): although the effect of losses is more pronounced (in the Bell and Bell+Non-resonant cases), it still remains limited to high energies, while no systematic difference is visible in the range $10-1000$ GeV.

\section{Discussion}
\label{sec:discuss}

SNRs contribute a spectrum of CRs that is made of two terms: one is the flux of particles that escape from upstream of the shock and the the other is the flux of particles that are liberated into the ISM when the shock is dissipated and the particle initially trapped downstream are free to leave the remnant. For protons, the escape flux, at any given time, is strongly peaked around the maximum energy at that time. The integration over the whole history results in a broad spectrum that roughly extends from the maximum momentum reached at the end of the ST phase and the maximum momentum reached at the beginning of the same phase. On the other hand, protons that are trapped in the downstream lose energy adiabatically and these losses reflect onto the normalisation and shape of the CR spectrum liberated into the ISM.

For electrons, the situation is somewhat more complex: electrons are sensitive to energy losses both through the maximum energy that they can achieve, and because of the radiative losses that they suffer while being advected downstream. Their escape from upstream is similar to that of protons when their maximum energy is not limited by losses, while drops to zero when radiative losses are the main limitation to electrons' acceleration. 

We describe both protons and electrons, and both their escape flux from upstream and at the end of the evolution of the SNR, for different types of SNRs. We calculate the magnetic field in the acceleration region using the excitation rate of the non-resonant hybrid instability induced by CR protons and its saturation level. The magnetic field produced by the streaming of CR protons also determines the maximum energy and the radiative losses of electrons. 

Our main conclusions can be summarised as follows:

1) In the context of non--resonant streaming instability, which is the mechanism expected to drive magnetic field amplification in the early stages of the evolution of SNRs, only very powerful and rare events, resulting from the explosion of energetic SNe (what we called type II* SNRs) can accelerate protons up to the PeV range. In these cases, a pronounced feature appears in the total spectrum around $\sim 100$ TeV, where the spectrum liberated at the end of the SNR evolution merges with the escape flux integrated over time. A weaker feature is present at $\sim 10$ TeV, because of the complex environment in which these  SN explosions take place. For type Ia SNe, the maximum energy of protons as derived in our calculations is $\sim 100$ TeV, but there is a pronounced dip in the spectrum at $\sim 10$ TeV. For type II SNe, the spectrum is strongly suppressed at energies of $\sim$ few tens of TeV. In all cases, at low energies the spectra are power laws. 

2) The maximum energies of protons as discussed above are mainly the result of the self-confinement of CRs in the shock proximity, due to the excitation of the current driven non-resonant streaming instability (see also~\citet{cristofari2020}). We calculated the strength of the magnetic field downstream of the shock, expected based on this process, as a function of the shock velocity, and compared the results with the observed trend, as reported by \cite{vink2012}. The scaling $B_{2}^{2}/\rho\propto v_{s}^{3}$, typical of this instability, is in good agreement with the sparse data available, and provides the correct normalisation. For old SNRs, with typical velocity $v_{s}\lesssim 1000$ km/s, the non-resonant instability is quenched and the strength of the magnetic field is comparable with that in the surrounding medium $\delta B\lesssim B_{0}$.  A recipe for magnetic field evolution often used in the literature was put forward by~\citet{caprioli11} and~\citet{morlino2012}. For typical speeds of SNR shocks, such a recipe leads to larger values of the amplified field, especially for older SNRs. As discussed in \S \ref{sec:Bfield}, this recipe has numerous caveats that it is important to keep in mind.

3) The amplified magnetic field described above plays a crucial role in shaping the spectrum of the accelerated electrons that are advected downstream, because of radiative losses.  In the cases we investigated, a substantial steepening of the spectrum was only obtained when the phenomenological prescription of~\citet{caprioli11} was adopted, that leads to results in qualitative agreement  with those of~\citet{rebecca}. When the magnetic field is assumed to be amplified according with the growth of the non resonant \citep{bell2004} and resonant streaming instability \citep{amatoblasi}, the effect of losses on the electron spectrum typically reduces to a cutoff in the overall spectrum rather than a broad steepening.

The appearance of a steepening in the electron spectrum (with respect to the proton spectrum) depends on the time dependence of the maximum electron energy $p^e_{\rm max}(t)$. In the initial stages of the SNR evolution (ejecta-dominated to early ST phase) the maximum energy is certainly determined by energy losses: in the few hundred $\mu G$ typical of a SNR at the beginning of the ST phase, synchrotron losses degrade the particle energy to $\sim 10$ GeV by the end of the remnant life. Later times (in the ST phase) may contribute electrons with increasingly larger energies if the amplified magnetic field decreases sufficiently fast with the shock velocity, as it is typically the case for all the prescriptions above. When $p^e_{\rm max}(t)$ becomes age-limited, as for protons, contributions from different times pile up, eventually leading to an overall cutoff in the TeV range. In summary, sustained magnetic field amplification may provide a steepening in the electron spectrum, but specific recipes for the amplification determine whether the steepening extends over several orders of magnitude in energy or is rather limited to a narrow energy range, thereby appearing as a cutoff in the electron spectrum. Current data suggest that a \emph{global} steepening must be present between $\sim 10$ GeV and a few TeV \citep{EvoliPRL}, although it is not clear that synchrotron losses are the cause of this phenomenon~\citep{ohira+12}. 


 This result shows that if SNRs are responsible for the observed spectrum of electrons, the late phases of their evolution, usually considered of little interest for cosmic ray acceleration, are in fact crucial, in that synchrotron energy losses may be at work for long times. On the other hand, the streaming instabilities that are thought to be most effective in amplifying the magnetic field may not be able to provide the magnetic fields that, through synchrotron losses, would cause the required steepening. Since such stages are crucial to shape the spectrum of electrons, it is worth asking whether there are physical phenomena that we did not take into account that might lead to larger (or smaller) fields at late times: one point to keep in mind is that the total compression factor at CR modified shocks might be somewhat larger than $4$ if CR acceleration remains efficient; for instance, in the simulations of \cite{haggerty+20} one has $r\sim 6-7$. This can indeed lead to somewhat larger magnetic fields, as one can infer from Eq. \ref{eq:B2}. However, in order to have a sizeable effect on the spectrum of electrons, this should happen for low shock velocity, while it is typically expected to be more of a concern for fast shocks. One could also speculate that, in addition to CR induced magnetic field amplification (that takes place upstream), the field could be further amplified downstream of the shock, perhaps through Richtmeier-Meshkov instability. 
 As discussed above, also the duration of the ST phase, which may be controlled also by the non-thermal cosmic ray pressure \cite[]{2018PhRvL.121i1101D}, may affect the spectrum of electrons. 
 Finally, we note that if shock acceleration were efficient even when the shock reaches Mach number of a few, the spectrum of accelerated particles, which is affected by the compression factor, would become steeper. For typical values of the parameters, when the shock reaches the radiative phase, its Mach number is still $\sim 40$, so it is possible that the CR spectrum may still evolve  during the early radiative stage. 
 
 

Clearly, there are also physical processes that lead to smaller magnetic fields than predicted above, thereby making the spectral modification for electrons even less effective. For instance, when the shock moves in the ordinary ISM, as it is typically the case in the late stages of the evolution of a SNR, neutral hydrogen induces a strong level of ion-neutral damping, that substantially limits the growth of perturbations. These effects are not included in our calculations, not in numerical simulations of DSA.  

From the discussion above, it is clear that if the difference in the spectrum of electrons and protons is to be attributed to radiative losses in the downstream of a SNR shock, the required conditions appear to be rather at odds with the ones that we would typically assume to exist. In particular, some rather efficient mechanism for magnetic field amplification should come into effect for late, slow moving shocks. A dedicated effort to investigate these stages is definitely needed.  

On the other hand, if to take the results of our investigation at face value, then the spectral shape of electrons and protons liberated into the ISM by an individual SNR should be very similar, hence it would follow that the observed difference should be attributed to phenomena occurring after the particles have been released into the ISM. If the diffusion coefficient describing transport in the Galaxy is the same for the two species, as one should expect, the only possibility left open is that electrons and protons may develop different spectral shapes while propagating in the neighbourhood of the source, due to the large perturbations induced by the escaping particles~\citep{schroer2020}. This possibility is currently being investigated.

\section*{Acknowledgments}
We thank Rebecca Diesing and Rino Bandiera for interesting discussions about electron spectra and radio emission from SNRs.The research of PB was partially funded through Grant ASI/INAF No. 2017-14- H.O.
DC was partially supported by NASA (grants NNX17AG30G, 80NSSC18K1218, and 80NSSC18K1726) and by NSF (grants AST-1714658, AST-1909778, and PHY-2010240).

\bibliographystyle{aa} 
\bibliography{SNR.bib} 

\end{document}